\documentclass[journal]{IEEEtran}
\usepackage{cite}
\usepackage{amsmath,amssymb,amsfonts}
\usepackage{algorithmic}
\usepackage{graphicx}
\usepackage{textcomp}
\usepackage[monochrome]{xcolor}
\usepackage{threeparttable}
\usepackage{multirow}
\ifCLASSINFOpdf
\else
\fi

\begin{document}
\title{Integrated Human Activity Sensing and Communications}

\author{Xinyu~Li,~\IEEEmembership{Student Member,~IEEE,}
Yuanhao~Cui,~\IEEEmembership{Member,~IEEE,}
J.~Andrew~Zhang,~\IEEEmembership{Senior Member,~IEEE,}
Fan~Liu,~\IEEEmembership{Member,~IEEE,}
Daqing Zhang,~\IEEEmembership{Fellow,~IEEE,}
Lajos Hanzo,~\IEEEmembership{Fellow,~IEEE}

\thanks{X. Li and J. A. Zhang are with the Faculty of Engineering and Information
Technology, University of Technology Sydney, Ultimo, NSW 2007, Australia (e-mail: Xinyu.Li-1@student.uts.edu.au; andrew.zhang@uts.edu.au).}

\thanks{Y. Cui is with the School of Information and Communication Engineering, Beijing University of Posts and Communications (BUPT), Beijing 100876, China (email:cuiyuanhao@bupt.edu.cn).}

\thanks{F. Liu is with the Department of 
Electronic and Electrical Engineering, Southern University of Science and Technology, Shenzhen 518055, China (e-mail: liuf6@sustech.edu.cn).}


\thanks{D. Zhang is with Telecom SudParis, France and Peking University, China. (e-mail: daqing.zhang@telecom-sudparis.eu).}

\thanks{L. Hanzo is with the School of Electronics and Computer Science, University of Southampton, Southampton SO17 1BJ, U.K. (e-mail:
lh@ecs.soton.ac.uk).}
}


\maketitle

\begin{abstract}

Advances in wireless communication and signal processing facilitate integrated sensing and communication (ISAC) - a compelling technology that intrinsically combines sensing and communication functionalities for the dual-purpose exploitation of wireless/hardware resources and pursues mutual benefits. Consequently, the next-generation communications network will be \textit{perceptive}. In this article, we provide a review of human-related sensing in the context of ISAC. We first present a general ISAC receiver signal processing framework, with a focus on human activity recognition (HAR). Based on its specific spatial deployments, we then categorize ISAC HAR into monostatic, bi-static, and distributed deployments, and discuss their properties, critical research problems and solutions. To facilitate the system's realization and improve its recognition performance, we then explore the inherent connections between the physical-layer system parameters and HAR performance metrics. Experimental results are presented for characterizing the sensing potentials of different ISAC systems. Finally, we review the technical challenges and identify the open research problems.
\end{abstract}


\IEEEpeerreviewmaketitle

\section{Introduction}
\IEEEPARstart{B}{oosted} by innovation in artificial intelligence and sensing technologies, human-centric decision-making systems are under intensive investigation in a broad range of compelling applications, such as safety protection, smart cities, and remote health-care. Given their ubiquitous, contactless and all-weather availability, the potential of standardized wireless signals (WiFi, LoRa, etc.) has been explored for identifying the nature of human motions, relying for example on the received signal strength indicator (RSSI), and channel state information (CSI) \cite{tan2018exploiting}. At the time of writing, these wireless sensing functionalities are primarily implemented by exploiting the reference signals of the standardized WiFi or fourth-generation (4G)/fifth-generation (5G) waveforms. Hence, the quality of the above sensory data is fundamentally determined by the pilot structure, the standardized waveform, and the spatial relationships of the commodity wireless transceivers deployed. However, these systems are primarily designed for communication, rather than for sensing. 

Human activities do influence the wireless signal propagation properties such as reflection, diffraction and scattering, which provide human activity sensing opportunities through analyzing and mapping the variations of the received signals to a specific activity \cite{ma2019wifi}. However, its recognition accuracy is subject to the constraints of the communication protocols and multipath effects. When complementing the communication functionality with built-in sensing functionality \cite{cui2021integrating}, we may appropriately apportion the spatial/time/frequency-domain wireless resources between sensing and communications more explicitly, and explore the hitherto untapped signal structures for sensing usage relying on data payloads, rather than only using the standardized pilot structure \cite{9737357}. On the other hand, compared to the contactless sensors embedded in the environment, ISAC has the potential of economically constructing an intelligent system, which is sensitive to the surrounding variations. Therefore, it is essential to critically appraise the potential of wireless communication to infer the nature of human activity from the raw data with the aid of sophisticated processing and recognition algorithms. This allows us to provide a bird's eye view of the new research in this area. 

\begin{figure*}[!t]
	\centering 
	\includegraphics[width=0.8\linewidth]{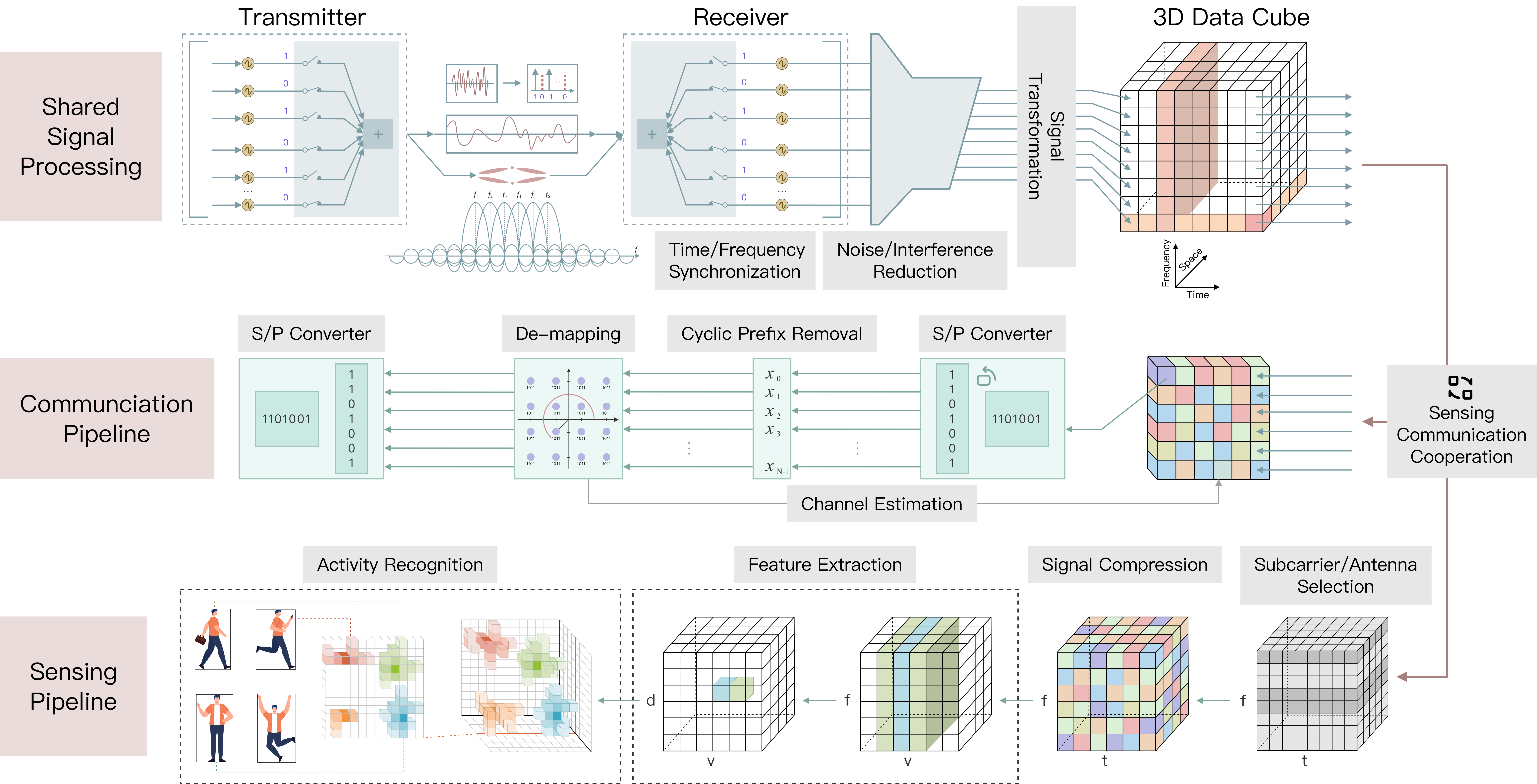}
	\caption{The general pipeline of HAR in ISAC.
	}
	\label{Fig1}
\end{figure*}

We commence by providing a review of human activity-related sensing in the context of ISAC. We first provide a systematic overview of ISAC signal processing framework, which elaborates on the connections and distinctions between the wireless communication and sensing processing pipelines. Then, to explore the impact of the wireless devices' geographical and spatial relationships with the sensing performance, we identify three typical ISAC configurations (e.g., monostatic, bistatic, and distributed configurations), followed by discussing their respective key challenges. Furthermore, we analyze the impact of several physical ISAC parameters on the HAR performance and discuss the relevant optimization principle by jointly considering human activity sensing and communications. Then, some experimental results are provided for illustrating the potential of different ISAC systems. Finally, we enlist some research challenges and open research opportunities before presenting concluding remarks.

\section{A Systematic View of ISAC Signal Processing}
\label{section2}

To provide a systematic view of the evolution from the communication-only devices to the ISAC infrastructure, in this section, we introduce a general ISAC receiver signal processing framework by examining the connections and distinctions of the operational communication and sensing signal processing procedures, with reference to the sensing application of HAR. For each procedure, various state-of-the-art technologies and the corresponding challenges are detailed.

\subsection{The Shared Procedures}

This subsection presents the shared reception procedures of the communication and sensing signal processing pipelines, as shown at the top of Fig.~\ref{Fig1}.

\subsubsection{Time/Frequency Synchronization}
Time/frequency synchronization is a fundamental requirement for ISAC systems to achieve both high communication data rates and sensing accuracy. Generally, the communication requirement can be satisfied by compensating for the offsets caused by clock asynchronism using embedded pilots, or by absorbing the offsets into channel estimation. However, for time-of-arrival (ToA)-based sensing, the residual offsets can still trigger estimation ambiguity and consequently produce ghost targets. Assume that the carrier frequency of an OFDM system is 3.5GHz, and the oscillator’s stability is 10 parts per million (ppm). Then, the carrier frequency offset (CFO) can be as high as 3.5 GHz $\times$ 10 ppm= 35 kHz. Even when 10 Hz residual CFO is left after a compensation algorithm dedicated to communications, 
the estimation error of human radial velocity still reaches 0.86 m/s. Therefore, sophisticated synchronization methods should be devised for ensuring high-accuracy HAR. 



\subsubsection{Noise/Interference Reduction}
Signal impairments such as interference constitute unintended but ubiquitous contamination of any radio system, and a low signal-to-interference-plus-noise ratio (SINR) severely degrades both the communication and sensing performance. However, the communication and sensing functionalities react differently to interference. For instance, all propagation paths contain valuable signals for communication, while some paths (e.g., the paths that are not reflected by the targets of interests)  are undesired for sensing and shall be treated as interference.

\begin{table*}[ht!]
\centering
\caption{Sensing characteristics of diverse wireless networks.}
\begin{threeparttable}
\begin{tabular}{|c|c|ccccccc|}
\hline
                                                                            &                                                                                                                    & \multicolumn{1}{c|}{802.11n}                                                                                & \multicolumn{1}{c|}{802.11ac}                                                                                                & \multicolumn{1}{c|}{802.11ad}                                          & \multicolumn{1}{c|}{802.11ax}                                                               & \multicolumn{1}{c|}{LTE}                                                                             & \multicolumn{1}{c|}{NR}                                                                                                                                       & LoRa                                                                          \\ \hline
\multirow{5}{*}{\begin{tabular}[c]{@{}c@{}}Signal\\ Structure\end{tabular}} & \begin{tabular}[c]{@{}c@{}}Sensing Range\\ Resolution\end{tabular}      & \multicolumn{1}{c|}{\begin{tabular}[c]{@{}c@{}}$\sim$2 m\\$\sim$5 m\end{tabular}}                          & \multicolumn{1}{c|}{\begin{tabular}[c]{@{}c@{}}$\sim$1 m\\$\sim$2 m\\$\sim$5 m\end{tabular}}                                  & \multicolumn{1}{c|}{$\sim$0.05 m}                                             & \multicolumn{1}{c|}{\begin{tabular}[c]{@{}c@{}}$\sim$1 m\\$\sim$2 m\\$\sim$5 m\end{tabular}} & \multicolumn{1}{c|}{\begin{tabular}[c]{@{}c@{}}$\sim$10 m\\$\sim$50 m\\$\sim$100 m\end{tabular}} & \multicolumn{1}{c|}{\begin{tabular}[c]{@{}c@{}}$\sim$0.5 m\\$\sim$5 m\\$\sim$30 m\end{tabular}} & \begin{tabular}[c]{@{}c@{}}$\sim$500 m\\$\sim$1000 m\end{tabular}                   \\ \cline{2-9} 
                                                                            & \begin{tabular}[c]{@{}c@{}} Signal for\\ Sensing$s^{1}$\end{tabular}         & \multicolumn{1}{c|}{\begin{tabular}[c]{@{}c@{}}STF\\ LTF\\ \end{tabular}} & \multicolumn{1}{c|}{\begin{tabular}[c]{@{}c@{}}STF\\ LTF\\ \end{tabular}} & \multicolumn{1}{c|}{\begin{tabular}[c]{@{}c@{}}STF\\ CEF\end{tabular}} & \multicolumn{1}{c|}{\begin{tabular}[c]{@{}c@{}}STF\\ LTF\end{tabular}}            & \multicolumn{1}{c|}{\begin{tabular}[c]{@{}c@{}}DMRS\\SRS, CRS\\CSI, PRS\end{tabular}}                    & \multicolumn{1}{c|}{\begin{tabular}[c]{@{}c@{}}DMRS\\PTRS\\ SRS, CSI\end{tabular}}                                                                       & \begin{tabular}[c]{@{}c@{}}Upchirps \\ Sync Word\\ Downchirps\end{tabular}    \\ \cline{2-9} 
                                                                            & \begin{tabular}[c]{@{}c@{}}Available\\ Frequency\end{tabular}          & \multicolumn{1}{c|}{\begin{tabular}[c]{@{}c@{}}2.4 GHz\\ 5 GHz\end{tabular}}                                  & \multicolumn{1}{c|}{5 GHz}                                                                                                    & \multicolumn{1}{c|}{60 GHz}                                             & \multicolumn{1}{c|}{\begin{tabular}[c]{@{}c@{}}2.4 GHz\\ 5 GHz\end{tabular}}                  & \multicolumn{1}{c|}{\begin{tabular}[c]{@{}c@{}}800 MHz \\ 1.8 GHz\\ 2.6 GHz\end{tabular}}             & \multicolumn{1}{c|}{\begin{tabular}[c]{@{}c@{}}7.125 GHz\\ 52.6 GHz\end{tabular}}                                                                               & \begin{tabular}[c]{@{}c@{}}169 MHz\\ 433 MHz\\ 868 MHz\\ 915 MHz\end{tabular} \\ \cline{2-9} 
                                                                            & \begin{tabular}[c]{@{}c@{}}Signal\\ Typ$e^{2}$\end{tabular}              & \multicolumn{3}{c|}{OFDM}                                                                                                                                                                                                                                                                                                                                                                                                                                                             & \multicolumn{1}{c|}{OFDMA}                                                                  & \multicolumn{1}{c|}{\begin{tabular}[c]{@{}c@{}}SC-FDMA\\ OFDMA\end{tabular}}                         & \multicolumn{1}{c|}{\begin{tabular}[c]{@{}c@{}}DFT-S-OFDM\\ CP-OFDM\end{tabular}}                                                                             & Chirps                                                                        \\ \hline
\multirow{4}{*}{Deployment}                                                 & Coverage                                                                                              & \multicolumn{1}{c|}{$\sim$8 m}                                                                              & \multicolumn{1}{c|}{$\sim$5 m}                                                                                               & \multicolumn{1}{c|}{$\sim$5 m}                                         & \multicolumn{1}{c|}{$\sim$8 m}                                                             & \multicolumn{1}{c|}{$\sim$15 m}                                                                     & \multicolumn{1}{c|}{/}                                                                                                                               & $\sim$50 m                                                                     \\ \cline{2-9} 
                                                                            & \begin{tabular}[c]{@{}c@{}}Cooperative\\ Sensing\end{tabular}      & \multicolumn{5}{c|}{\begin{tabular}[c]{@{}c@{}}Protocol \\  supported\end{tabular}}                                          & \multicolumn{1}{c|}{\begin{tabular}[c]{@{}c@{}}Protocol\\ supported\end{tabular}}                                                                             & \begin{tabular}[c]{@{}c@{}}Protocol \\ not supported\end{tabular}             \\ \cline{2-9} 
                                                                            & \begin{tabular}[c]{@{}c@{}}Data\\ Fusion\end{tabular}              & \multicolumn{4}{c|}{\begin{tabular}[c]{@{}c@{}}Decentralized fusion\\ Non-cooperative fusion\end{tabular}}                                                                                                                                                                                                                                                                                                                                                                                                                                                                          & \multicolumn{3}{c|}{\begin{tabular}[c]{@{}c@{}}Centralized/Decentralized fusion\\ Cooperative/Non-cooperative fusion\end{tabular}}                                                                                                                                                                                                                   \\ \hline
\multirow{3}{*}{\begin{tabular}[c]{@{}c@{}}Data \\ Processing\end{tabular}} & \begin{tabular}[c]{@{}c@{}}Computing\\ Hardware\end{tabular}       & \multicolumn{4}{c|}{\begin{tabular}[c]{@{}c@{}}Access point\\ Phone\end{tabular}}                                                                                                                     & \multicolumn{1}{c|}{\begin{tabular}[c]{@{}c@{}}BS\\ Phone\end{tabular}}                    & \multicolumn{2}{c|}{\begin{tabular}[c]{@{}c@{}}BS\\ Phone\\ Edge device\end{tabular}}                                                                                                                                             \\ \cline{2-9}                                                                             & \begin{tabular}[c]{@{}c@{}}Measurement\\ for HAR\end{tabular}      & \multicolumn{4}{c|}{\begin{tabular}[c]{@{}c@{}}CSI, RSSI \\ Round trip time (RTT)\end{tabular}}                                                                                                                                                                                                                                                                                                                                                                                                                                                                                      & \multicolumn{1}{c|}{\begin{tabular}[c]{@{}c@{}}CSI\\ RSS\end{tabular}}                               & \multicolumn{1}{c|}{\begin{tabular}[c]{@{}c@{}}RSSI, CSI\\ Amplitude\\ Phase\end{tabular}}                                                                    & \begin{tabular}[c]{@{}c@{}}Amplitude\\ Phase\end{tabular}                     \\ \cline{2-9}                                                                             & \begin{tabular}[c]{@{}c@{}}HAR Algorithm\\ Challenge\end{tabular}  & \multicolumn{4}{c|}{\begin{tabular}[c]{@{}c@{}}HAR with small-size data\\ Generalization to new persons\\Location/orientation-dependent HAR\end{tabular}} & \multicolumn{3}{c|}{\begin{tabular}[c]{@{}c@{}}Location/orientation-dependent HAR\\Outdoor long-range interference\\Recognition of remote micro-activities\\Slow fading\end{tabular}}   \\                                                                                                                                                    \hline

\end{tabular}

      \begin{tablenotes}
        \footnotesize
         \item[1]SFD: Start Frame Delimiter; STF: Short Training Field; LTF: Long Training Field; CEF: Channel Estimation Field; DMRS: Demodulation Reference Signal; SRS: Sounding Reference Signal; CRS: Cell Reference Signal; CSI: Channel State Information; PRS: Positioning Reference Signals; PTRS: Phase Tracking Reference Signal. ${}^{2}$OFDMA: Orthogonal Frequency Division Multiple Access; SC-FDMA: Single Carrier-FDMA; DFT-S-OFDM: Discrete Fourier Transform-Spread OFDM; CP-OFDM: Cyclic Prefix-OFDM.
      \end{tablenotes}
      \end{threeparttable}
\label{t3}
\end{table*}

\subsubsection{Signal Transformation} 
 After time-domain pre-processing, the ISAC signal is subjected to space-time-frequency analysis. For instance, the time-domain (TD) signals can be converted to the frequency domain (FD) with the fast Fourier transform (FFT), and the attained Doppler frequency can be used for inferring human activities. A typical signal transformation is shown in Fig.~\ref{Fig1}, which transforms the time series of received signals into a complex-valued 3-dimensional (3D) data cube. The data cube characterizes both the information-carried communication symbols and the channel variations in time, frequency, and spatial domains, and hence can be shared between the communication and sensing functionalities. 


\subsubsection{Signal Separation}
It is vitally important to be able to distinguish sensing echoes from the received multipath communications signals for the subsequent HAR procedure, which remains an open issue at the time of writing.

 

\subsection{Separate Procedures for Sensing}
In this subsection, we mainly focus on the signal processing procedure of the sensing pipeline. Additionally, we summarize the sensing characteristics of several existing wireless networks from their signal structure, network deployment, and data processing perspectives, as shown in Table~\ref{t3}.

\subsubsection{Subcarrier/Antenna Selection}
Due to the frequency-selective fading and antenna topology, the variation patterns of human echoes associated with different subcarriers/antennas may be diverse, and are susceptible to factors such as the moving direction of human targets. The received sensing signals that exhibit a weak response to the human target movement cannot improve the HAR performance. In this case, harnessing subcarrier/antenna selection is indispensable for capturing the signals that show significant fluctuations with human movement \cite{wang2017device}.

\subsubsection{Signal Compression}
The goal of signal compression is to remove the redundancy in the discrete 3D range-Doppler-angle sensing signals, such as static background clutter and outdoor signal contamination. There are two main signal compression strategies: statistical dimension reduction approaches and clustering approaches based on the range-Doppler-angle estimates. 
However, since each dimension in the 3D data cube has a clear physical meaning, dimension reduction techniques may destroy the structure of the data cube and prevent the physical interpretation of the movements within the cube. By contrast, the clustering-based strategy is capable of retaining the 3D data structure \cite{9194046}, and the sensing data of the human target can be represented as a 3D range-Doppler-angle point cloud for the subsequent feature extraction.

\subsubsection{Feature Extraction}
Radio features may be extracted by manual feature engineering, or by automatic deep learning (DL) algorithms. In manual feature engineering, the amplitude and phase of the received sensing signals are commonly used features, because they can characterize the impact of human activity on signal propagation. Additionally, the time-varying Doppler/micro-Doppler frequency shifts \cite{9194046} contained in signal phases, which correspond to radial velocities of different components of the human body, are also effective for single-person activity recognition. Additionally, in multi-person scenarios, spatial parameters such as the range and angle are indispensable for separating the signals reflected by different targets. When using cloud of points for feature extraction, the intensity of the reflected points and the shape of the point cloud can also be adopted for characterizing diverse human activities. Again, DL \cite{8067693} constitutes an effective tool for automatically extracting HAR-related features. 
Early research has exploited the combination of convolutional neural networks (CNNs) with radio-based HAR feature extraction. However, the temporal information of human motions gradually vanishes during CNN training, which may gravely degrade the classification performance in the next stage. Hence, despite their higher computational complexities, the memory-enabled recurrent NNs are routinely adopted for extracting time-varying HAR-related features, resulting in significantly improved performance. 



\subsubsection{Activity Recognition}
The HAR algorithms may be divided into two categories: model-based and learning-based algorithms \cite{8067692}. Model-based approaches, such as the Fresnel Zone model and the Angle of Arrival model, mathematically characterize the underlying relationship between human motion and the resultant signal variations. Hence the human movement-related parameters may be quantified with the aid of the signal dynamics. Given their clear physical interpretations, model-based algorithms have great potential in accomplishing fine-grained activity recognition tasks, and can support the exploration of sensing limits (e.g., sensing coverage and performance bound) of HAR. On the other hand, learning-based approaches \cite{8067692}, including machine learning (ML)-based and DL-based techniques \cite{9194046,10.1145/3411816,8010417}, aim for learning the mapping between sensing measurements and the label of the corresponding human activities by using pre-extracted features. Generally, both manually and automatically extracted features can be employed in learning-based methods. However, in most cases, the DL-based feature extractor and classifier are combined to process the input data and then, to classify human activities in an end-to-end manner without any human intervention. 




\section{Unique Signal Processing for HAR in Practical ISAC Deployments}
\begin{figure}[!t]
	\centering 
	\includegraphics[width=\columnwidth]{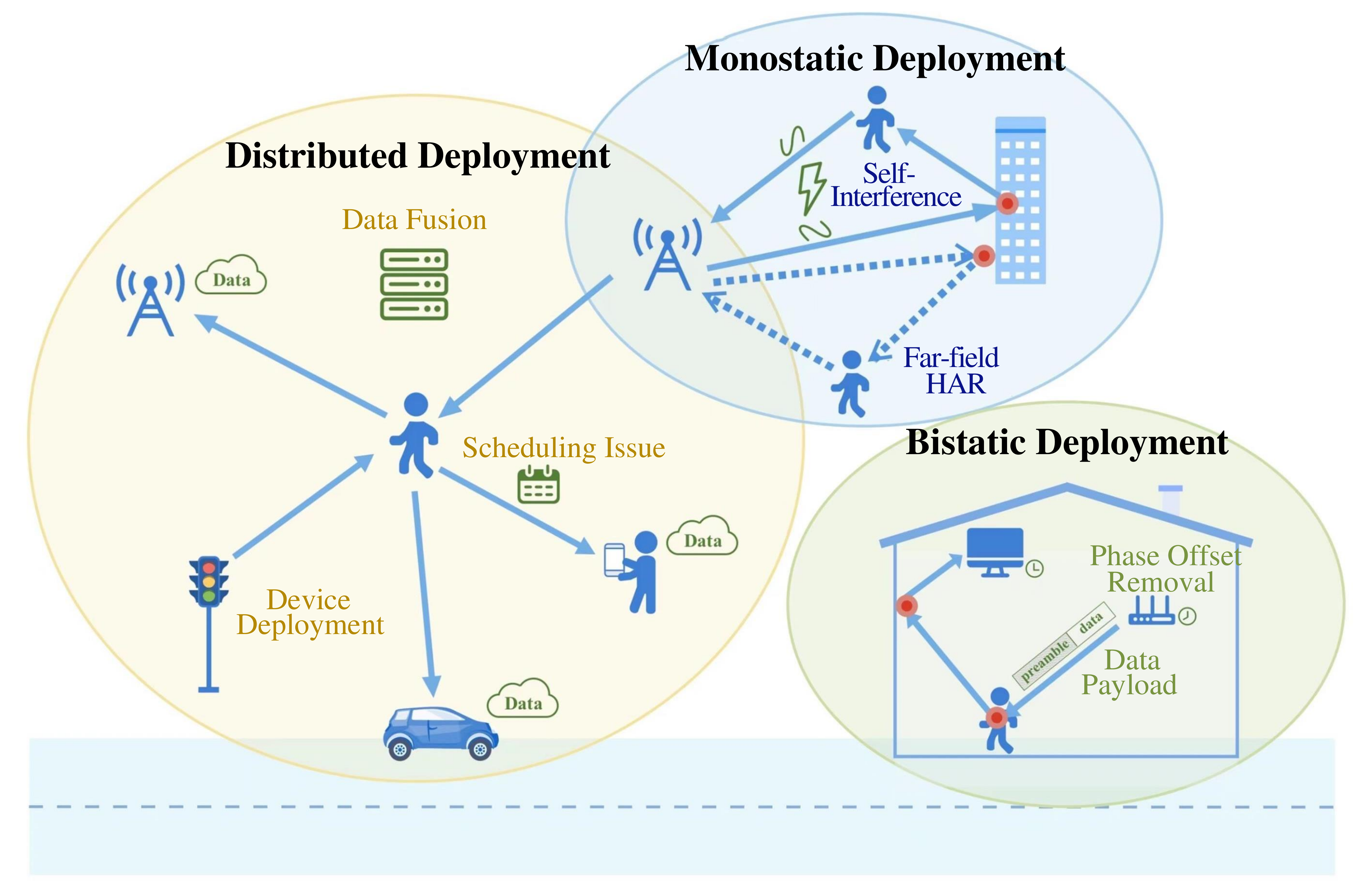}
	\caption{Three deployments of ISAC systems for HAR. }
	\label{Fig2}
\end{figure}

\
\label{section3}
There are three main deployments of ISAC systems for HAR, namely monostatic, bistatic, and distributed deployments (see Fig. \ref{Fig2}), depending on the locations of transmitters and receivers. In this section, we discuss the unique properties of these deployments beyond the general signal processing operations, and present their key problems when used for HAR.

\subsection{Monostatic Deployment}
The monostatic system in Fig. \ref{Fig2} transmits wireless signals to sense the environment and captures the target echoes via the sensing receiver co-located and synchronized with the transmitter. For example, a 5G New Radio (NR) base station (BS) may sense the environment using the echoes of its transmitted downlink communication signals. Self-interference (SI) and far-field HAR are major concerns in such a deployment.

\subsubsection{Self-Interference (SI)}
In a monostatic system, the leaked out-of-band (OOB) transmitted signals generally interfere with the desired echo signals, also known as SI. The strong SI can saturate the receivers and overwhelm the target echoes. Since most modern communication systems dispense with time division multiple address (TDMA)-bursts and transmit continuous waveform, it is infeasible to use an idle period for receiving echoes after a TDMA-burstor as in a pulse radar. It is also impractical to adopt the transmitted signal as the local oscillator input for removing the SI, like in a frequency modulated continuous radar. Full duplex operation constitutes a long-term solution to this issue, as described in \cite{zhang2021enabling}, together with suboptimal near-term solutions, such as deploying a dedicated receiver antenna for sensing, widely separated from other antennas.

\subsubsection{Far-Field HAR}
According to the spatial relationship of the transmitters and targets in Fig. \ref{Fig2}, the sensing area may be divided into two distinct regions: near-field and far-field. 
In far-field, due to the low SINR, the signals reflected by the human target may be overwhelmed by the background clutter
\cite{7857052}. Explicitly, the micro-Doppler frequencies, which are eminently suitable for HAR, may become too weak to be captured. Alternatively, 
other features such as the time-varying range information of different human body segments can be employed for recognition \cite{8010417}.
\subsection{Bistatic Deployment}
Bistatic deployment refers to an ISAC system where the transmitter and receiver are spatially well-separated. A typical example is Wi-Fi sensing \cite{wang2017device}. Compared to monostatic systems, the bistatic deployment is more compatible with existing communication networks such as WiFi and cellular networks. Furthermore, the SI issue is naturally avoided by the spatially bistatic scheme. Nevertheless, phase offset removal and unknown data payload constitute pressing problems in bistatic deployments designed for accurate HAR.

\subsubsection{Phase Offset Removal}
Due to oscillator instability, phase offset tends to persist in the received signals of bistatic systems, leading to measurement ambiguity and accuracy degradation. For instance, sampling frequency offset (SFO) generally introduces violent fluctuations in the phase of sensing signals, and can even drown out the small phase changes caused by human movement. 
To compensate for the phase offsets and recover the information loss, cross-antenna correlation and cross-antenna ratio techniques \cite{10.1145/3411816,10.1145/3351279} may be applied by exploiting the fact that the phase offsets of different receiver antennas are the same. An alternative strategy discards the phase information and only uses the signal magnitude. But this results in degraded sensing performance \cite{ma2019wifi}. 

\subsubsection{Exploiting Data Payload for Sensing}
Operational communication-centric ISAC systems mainly utilize the known pilot signals for sensing applications. However, instead of only using a pilot, employing the entire frame as sensing signals may potentially achieve a higher signal-to-noise-ratio (SNR) and a finer Doppler frequency resolution. In most existing bistatic ISAC systems, the data payload is unknown at the receiver side. To this end, a possible strategy is to first decode the unknown data payload based on the channel estimation results, and then employ the entire frame for sensing. However, the quantitative benefit of achieving improved SNRs by the sensing-after-decoding scheme is yet to be verified, and may only be prominent over a limited range of SNRs.

\begin{table*}[]
	\centering
	\caption{Impact of ISAC Physical Parameters on HAR Performance.}
	\renewcommand\arraystretch{1.5}
	\begin{threeparttable}
		\begin{tabular}{|p{3.2cm}|p{5cm}|p{9cm}|}
			\hline
			\textbf{Physical Parameters}& \textbf{Impact on Sensing Pipeline} & \textbf{Impact on HAR performance} \\
			\hline 
			\textbf{\begin{tabular}[c]{@{}l@{}}Total Signal Bandwidth $B$\end{tabular}} & \begin{tabular}[c]{@{}l@{}}Larger $B$ leads to better range resolution \end{tabular}& \begin{tabular}[c]{@{}l@{}}Better multi-targets separation ability along range domain with larger $B$  \end{tabular} \\
			\hline
			\textbf{Carrier Frequency $f_c$} & \begin{tabular}[c]{@{}l@{}}Greater $f_c$ leads to finer velocity resolution \\ but smaller unambiguous velocity$^{2}$\end{tabular} & \begin{tabular}[c]{@{}l@{}} Different moving components of the human target can be be recorded with\\ finer granularity in velocity domain\end{tabular}\\
			\hline
			\begin{tabular}[c]{@{}l@{}}\textbf{Symbol} \textbf{Duration $T_{s}$}\end{tabular} & \begin{tabular}[c]{@{}l@{}}Larger $T_{s}$ leads to longer unambiguous\\range but lower range resolution\end{tabular}& \begin{tabular}[c]{@{}l@{}} 
				Longer detectable range with $T_s$ increasing, promising for far-field HAR\end{tabular}\\
			\hline
			\begin{tabular}[c]{@{}l@{}}\textbf{Subcarrier} \textbf{Interval $T$$^{1}$}\end{tabular} & \begin{tabular}[c]{@{}l@{}}Smaller $T$ leads to larger unambiguous\\velocity but lower velocity resolution\end{tabular}&\begin{tabular}[c]{@{}l@{}}Wider coverage in velocity domain to record the activities with higher velocity\\ components\end{tabular}\\
			
			\hline
			\textbf{Antenna Array Design} & \begin{tabular}[c]{@{}l@{}}Better antenna aperture and higher \\angular resolution with more antennas\end{tabular} &\begin{tabular}[c]{@{}l@{}} Locating human target more precisely at the angular direction \\ Distinguishing multiple targets at close angular directions\end{tabular}  \\
			\hline\textbf{Transmission Power} & \begin{tabular}[c]{@{}l@{}}Positively correlated with the \\coverage of the sensing system\end{tabular}  & \begin{tabular}[c]{@{}l@{}} Wider HAR coverage with higher transmission power \\ Stronger echo signals and more robust to interference\end{tabular}        \\
			\hline
			\textbf{Antenna Gain}  & \begin{tabular}[c]{@{}l@{}}Greater coverage in a certain\\ direction with higher antenna gain\end{tabular} & \begin{tabular}[c]{@{}l@{}} Wider HAR coverage \\Improving intensity of the reflected echoes with higher antenna gain \end{tabular}\\
			\hline
		\end{tabular}
		\begin{tablenotes}
			\footnotesize
			\item[1] For single-subcarrier signals, $T$ is $T_s$, and is euqal to 1/$B$.\\For OFDM signals, $T$ includes $T_s$ and cyclic prefix, and is equal to $N$/$B$, where $N$ is the number of subcarriers.
			\item[2] Maximum unambiguous velocity $v_{max}$ = $c/(2f_cT)$, and velocity resolution $\Delta v$ = $v_{max}$/$M$, where $c$ is the velocity of light, and $M$ is the number of symbols.
		\end{tablenotes}
	\end{threeparttable}
	\label{t1}
\end{table*}

\subsection{Distributed Deployment}
\label{section3-1}
In a distributed system, all transmitting and receiving devices are distributed in different spatial locations, which can provide spatial diversity gains for the illuminated target, yielding augmented human activity information to deal with target fluctuation \cite{zhang2021enabling}. In addition to the issues of the bistatic system, in a distributed system, conceiving a systematic design and topological arrangement, as well as sophisticated data fusion, infrastructure deployment, and scheduling are indispensable.

\subsubsection{Data Fusion}
Again, efficient data fusion is essential for removing data redundancy from the sources for accurate global feature representations. Explicitly, data-level, feature-level and decision-level fusion constitutes typical data fusion techniques \cite{JAVADI202048}. In data-level fusion, the raw data gleaned from different nodes are sent to the fusion center (FC) and then aggregated for extracting the HAR-related information. However, sending a huge amount of sensing data to the FC imposes a high communication burden and high hardware costs. By contrast, in feature-level and decision-level fusion, each node can preprocess its data and send the output features/decisions to the FC. This requires less data exchange between the FC and local nodes, saving both energy and computing resources at the FC. Furthermore, such decentralized strategies allow flexible algorithmic designs at different branches and hence can extract unique information from the devices.

\subsubsection{Deployment of Host and Slave Devices}
In a communication-centric ISAC system, both the host and the slave devices may act as sensing transceivers. However, host device deployments in pure communication systems and in ISAC systems generally follow different principles. For instance, in a cellular network, BSs are positioned to minimize interference between cells, which is not suitable for ISAC systems, because the interference also contains valuable sensing information. Therefore, in ISAC systems, the deployment of host devices faces a trade-off between communication interference and sensing performance. On the other hand, although slave device deployment is independent of the communication performance, the placement of terminals directly affects the coverage, orientation, and angles. Hence its placement has to be optimized for improved sensing performance.

\subsubsection{Scheduling Issue with Target Echoes}
In addition to data fusion and device deployment, sensing human activities also imposes challenges on the resource scheduling of distributed ISAC systems. Since the human echoes could randomly appear in the time, frequency, and spatial domains, novel ISAC scheduling algorithms are required for predicting the appearance of random echo signals and classifying the echoes in an orderly manner \cite{9737357}. Furthermore, intelligent resource allocation algorithms have to be designed for all distributed devices, based on the context information, such as the specific quality-of-service (QoS) requirements and battery consumption.

\subsection{Summary and Design Factors}


In light of the above discussions, it can be inferred that the distributed deployment yields the best HAR performance due to its spatial diversity gains and wide coverage, and has greater potential for through-the-wall HAR and compound activity recognition. However, more complex signal processing is required. In real-world applications, one can design bespoke ISAC deployment by considering the site and resource constraints.

In addition to their deployment, the system parameters of ISAC systems, as inherited from communication designs, may also have notable impact on HAR. In Table \ref{t1}, we present the impact of some salient physical ISAC parameters on HAR performance.

\section{Over-the-air Experiments and Results}
\begin{figure*}[h]
	\centering 
	\includegraphics[width=0.78\linewidth]{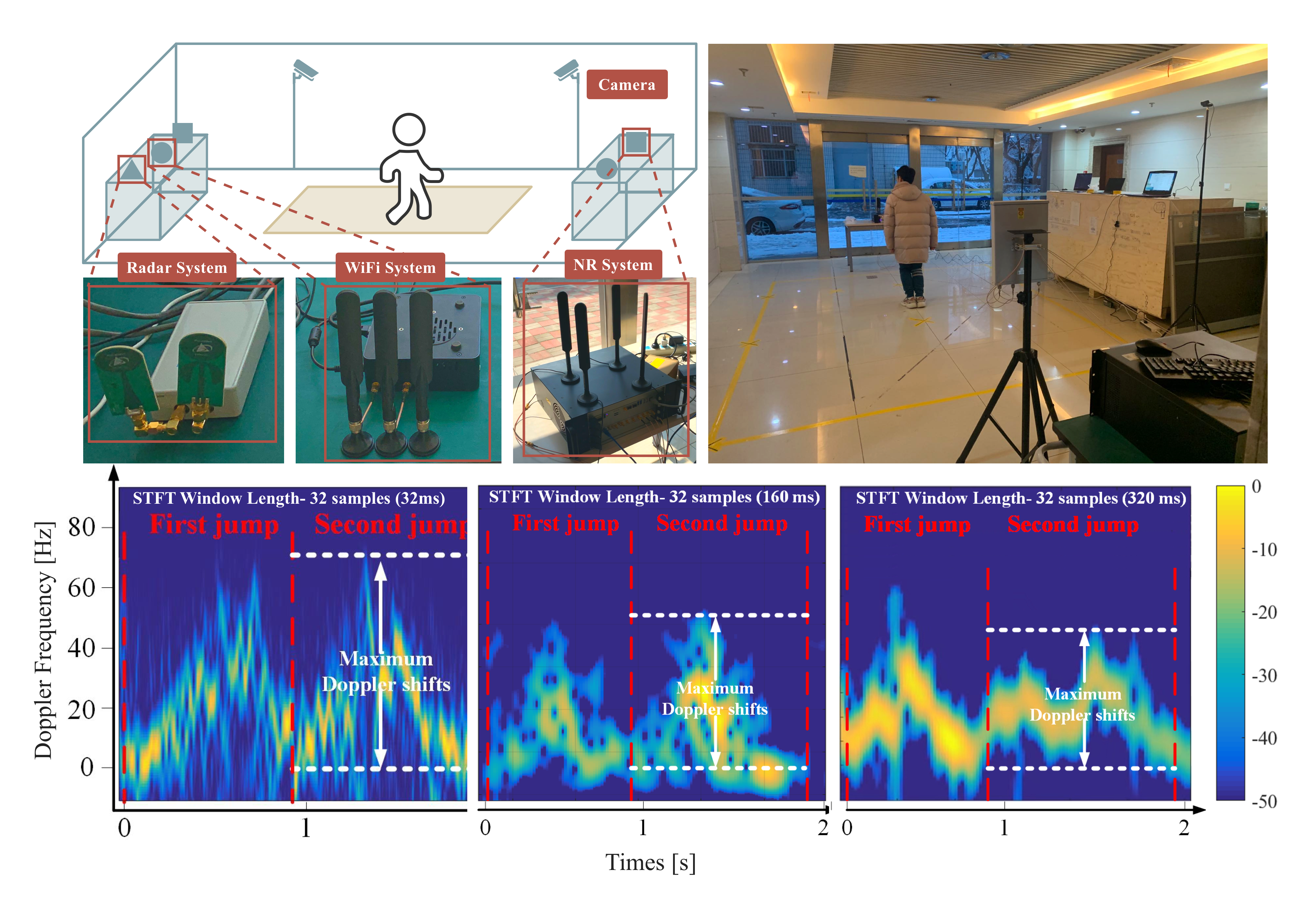}
	\caption{Experimental setup (top) and the resulting time-Doppler frequency spectrograms (bottom) of a human target jumping forward twice, for WiFi (left), radar (middle), and 5G NR (right) systems, respectively.
	}
	\label{Fig3}
\end{figure*}

In this section, we show some experimental results for different communication signals in HAR.

In the experiments, the reflected sensing signals corresponding to a human jumping forward in an indoor scenario of 20 m$^2$ were collected by three wireless systems: 1) a ultra-wideband (UWB) radar with 1.0 GHz bandwidth, 1.0 ms pulse repetition interval (PRI), and 4.0 GHz central frequency, 2) a WiFi system with 40 MHz bandwidth, 5.0 ms PRI, and 5.8 GHz central frequency, and 3) a 5G NR BS system with 100 MHz bandwidth, 10.0 ms PRI, and 3.6 GHz central frequency. Note that both the WiFi and NR systems collect signal reflections from separately deployed receiver antennas, so they are working in the bi-static sensing mode.
We preprocessed these sensing measurements using some of the methods described in Section \ref{section2} and \ref{section3}, and transformed the measurements into time-Doppler frequency spectrograms by using the short-time Fourier transform (STFT) with a window of 32 samples. 

To compare the sensing performances, we plot their micro-Doppler signatures in Fig. \ref{Fig3}. Observe that although the radial velocity components are of similar magnitude, the Doppler shifts produced by the system with a higher center frequency are more pronounced. More distinct frequency shift components facilitate the activity-related micro-Doppler features to be extracted from the reflected echoes manually, hence improving the HAR performance. Meanwhile, low PRI leads to finer time resolution, hence enabling the echo signals to convey the time-varying characteristics of human activities in more detail. Therefore, a system having low PRI can be used for distinguishing some similar human activities. Furthermore, the intensity of the NR spectrogram is the strongest, indicating that the NR system can be more robust to interference and has a wider coverage for HAR.  

\section{Challenges and Future Opportunities}
There exist a number of open problems in the research and development of HAR using ISAC signals. In this section, we
briefly discuss several research challenges and future opportunities.

\subsection{Multi-person Sensing}
When there are multiple moving persons in the sensing area, identifying the target of interest or recognizing the activities of multiple persons using wireless signals is a challenging problem. The general idea is to extract the reflected signals of each person and then recognize the corresponding human activity relying on the separated signals. This can be typically realized in two ways: separation via physical location and velocity \cite{9194046}, or separation via signal statistics \cite{10.1145/3411816}. In the first strategy, the signals for different targets may be separated from the spatial dimension by using the range, angle, and/or moving speed information, which requires high resolutions in these domains. In the second strategy, signals from multiple humans may be modeled as a linear sum of statistical independent signals, and the separation can be cast as a blind source separation problem.



\subsection{Joint Design and Optimization}

Communication and HAR generally have conflicting requirements for antenna placement and grouping. When using an array, sensing aims for increasing the antenna aperture and resolution by optimizing the element placements and virtual subarrays. By contrast, communication aims for high beamforming gain, for spatial diversity and for spatial multiplexing relying on a low signal correlation among antennas. Hence their joint design is challenging in this content. Considering the benefits of antenna grouping in communication and sensing, using hybrid antenna arrays is a low-cost, balanced option. 

Moreover, to achieve higher-accuracy HAR, the transmitted signals can be optimized by jointly considering several performance metrics for communication and HAR. An overview of such joint optimization for general ISAC scenarios is available in \cite{zhang2021enabling}. A promising avenue is to use multi-component Pareto-optimization for finding the full set of all optimal solutions, where none of the metrics can be improved without degrading at least one of the others. For HAR, signal optimization can be conducted by referring to the impacts presented in Table \ref{t1}. 

\subsection{Through-the-wall HAR}
Sensing through walls is also a challenging but compelling task in HAR. RF signals generally experience unpredictable reflection and absorption as they pass through walls, substantially weakening the received signals, hence reducing the HAR information. Furthermore, the characteristics of human activities buried in the received signals may become overwhelmed by environmental noise, affecting the subsequent feature extraction, especially upon distinguishing fine differences. In this case, approaches that can classify human activities behind the walls with minimal signal processing and human intervention are required \cite{8735849}. On the other hand, model-based algorithms may be conceived for characterizing the mathematical relationship between the received sensing signals and the human activities in non-line-of-sight through-the-wall scenarios.




\subsection{HAR Robustness and Generalization}
ISAC-aided HAR is sensitive to numerous factors, such as the sensing environment, network settings, relative location of human targets, geometry, and mobility situations. For instance, since different moving directions and orientations of a person with respect to the transceivers may result in various Doppler/micro-Doppler frequencies, how to improve the system's capability of recognizing human activities from diverse directions is a challenging issue. A feasible solution is to apply the Fresnel model \cite{8067692}. Additionally, due to the unique behavior of each individual,
it is essential to generalize the trained HAR algorithms, when new persons or new environments emerge.

\subsection{Distributed Sensing}
Given the potentially significant improvement in coverage and HAR accuracy, sensing based on a distributed topology is expected to be a general trend. However, research on sensing using off-the-shelf wireless devices under a distributed topology is still very limited. As presented in Section \ref{section3-1}, the challenges for distributed HAR mainly lie in the deployment and cooperation between transceivers, the fusion of data from diverse receivers, and the scheduling of the distributed devices. In addition, there is almost no discussion on the HAR performance bound of distributed sensing networks, which is also a promising research direction.

\section{Conclusions}
A review of human activity sensing in the context of ISAC was provided. We illustrated the general pipeline of ISAC signal processing for HAR and analyzed the sensing characteristics of several typical wireless networks. According to the spatial locations of transceivers, we categorized ISAC systems into three typical deployments and then, elaborated on characteristics and design strategies for HAR in these deployments. Furthermore, the impact of several physical ISAC factors on the HAR performance was discussed. Some experimental results were also provided to illustrate the potential and capabilities of different ISAC systems. Finally, we presented five key challenges and open research problems to facilitate a transition of the techniques into real-world applications.


%

\section*{Acknowledgements}

The authors would like to thank Prof. Octavia A. Dobre for the good discussions.


\ifCLASSOPTIONcaptionsoff
  \newpage
\fi



%
\bibliographystyle{IEEEtran}
\bibliography{reference}

\begin{thebibliography}{10}
\providecommand{\url}[1]{#1}
\csname url@samestyle\endcsname
\providecommand{\newblock}{\relax}
\providecommand{\bibinfo}[2]{#2}
\providecommand{\BIBentrySTDinterwordspacing}{\spaceskip=0pt\relax}
\providecommand{\BIBentryALTinterwordstretchfactor}{4}
\providecommand{\BIBentryALTinterwordspacing}{\spaceskip=\fontdimen2\font plus
\BIBentryALTinterwordstretchfactor\fontdimen3\font minus
  \fontdimen4\font\relax}
\providecommand{\BIBforeignlanguage}[2]{{%
\expandafter\ifx\csname l@#1\endcsname\relax
\typeout{** WARNING: IEEEtran.bst: No hyphenation pattern has been}%
\typeout{** loaded for the language `#1'. Using the pattern for}%
\typeout{** the default language instead.}%
\else
\language=\csname l@#1\endcsname
\fi
#2}}
\providecommand{\BIBdecl}{\relax}
\BIBdecl

\bibitem{tan2018exploiting}
B.~Tan, Q.~Chen, K.~Chetty, K.~Woodbridge, W.~Li, and R.~Piechocki,
  ``Exploiting {WiFi} channel state information for residential healthcare
  informatics,'' \emph{IEEE Communications Magazine}, vol.~56, no.~5, pp.
  130--137, May 2018.

\bibitem{ma2019wifi}
Y.~Ma, G.~Zhou, and S.~Wang, ``Wifi sensing with channel state information: A
  survey,'' \emph{ACM Comput. Surv.}, vol.~52, no.~3, May 2020.

\bibitem{cui2021integrating}
Y.~Cui, F.~Liu, X.~Jing, and J.~Mu, ``Integrating sensing and communications
  for ubiquitous {IoT}: Applications, trends, and challenges,'' \emph{IEEE
  Network}, vol.~35, no.~5, pp. 158--167, Sep. 2021.

\bibitem{9737357}
F.~Liu, Y.~Cui, C.~Masouros, J.~Xu, T.~X. Han, Y.~C. Eldar, and S.~Buzzi,
  ``Integrated sensing and communications: Towards dual-functional wireless
  networks for {6G} and beyond,'' \emph{IEEE Journal on Selected Areas in
  Communications}, pp. 1--1, 2022.

\bibitem{wang2017device}
W.~Wang, A.~X. Liu, M.~Shahzad, K.~Ling, and S.~Lu, ``Device-free human
  activity recognition using commercial {WiFi} devices,'' \emph{IEEE Journal on
  Selected Areas in Communications}, vol.~35, no.~5, pp. 1118--1131, May 2017.

\bibitem{9194046}
J.~Pegoraro, F.~Meneghello, and M.~Rossi, ``Multiperson continuous tracking and
  identification from mm-wave micro-{Doppler} signatures,'' \emph{IEEE
  Transactions on Geoscience and Remote Sensing}, vol.~59, no.~4, pp.
  2994--3009, April 2021.

\bibitem{8067693}
S.~Yousefi, H.~Narui, S.~Dayal, S.~Ermon, and S.~Valaee, ``A survey on behavior
  recognition using {WiFi} channel state information,'' \emph{IEEE
  Communications Magazine}, vol.~55, no.~10, pp. 98--104, 2017.

\bibitem{8067692}
D.~Wu, D.~Zhang, C.~Xu, H.~Wang, and X.~Li, ``Device-free {WiFi} human sensing:
  From pattern-based to model-based approaches,'' \emph{IEEE Communications
  Magazine}, vol.~55, no.~10, pp. 91--97, 2017.

\bibitem{10.1145/3411816}
Y.~Zeng, D.~Wu, J.~Xiong, J.~Liu, Z.~Liu, and D.~Zhang, ``Multisense: Enabling
  multi-person respiration sensing with commodity {WiFi},'' \emph{Proc. ACM
  Interact. Mob. Wearable Ubiquitous Technol.}, vol.~4, no.~3, Sep. 2020.

\bibitem{8010417}
B.~Jokanović and M.~Amin, ``Fall detection using deep learning in
  range-{Doppler} radars,'' \emph{IEEE Transactions on Aerospace and Electronic
  Systems}, vol.~54, no.~1, pp. 180--189, Feb. 2018.

\bibitem{zhang2021enabling}
J.~A. Zhang, M.~L. Rahman, K.~Wu, X.~Huang, Y.~J. Guo, S.~Chen, and J.~Yuan,
  ``Enabling joint communication and radar sensing in mobile networks - a
  survey,'' \emph{IEEE Communications Surveys \& Tutorials}, early access,
  2021.

\bibitem{7857052}
D.~Bleh, M.~Rösch, M.~Kuri, A.~Dyck, A.~Tessmann, A.~Leuther, S.~Wagner,
  B.~Weismann-Thaden, H.-P. Stulz, M.~Zink, M.~Rießle, R.~Sommer, J.~Wilcke,
  M.~Schlechtweg, B.~Yang, and O.~Ambacher, ``${W}$ -band time-domain
  multiplexing {FMCW} {MIMO} radar for far-field {3-D} imaging,'' \emph{IEEE
  Transactions on Microwave Theory and Techniques}, vol.~65, no.~9, pp.
  3474--3484, 2017.

\bibitem{10.1145/3351279}
Y.~Zeng, D.~Wu, J.~Xiong, E.~Yi, R.~Gao, and D.~Zhang, ``Farsense: Pushing the
  range limit of wifi-based respiration sensing with csi ratio of two
  antennas,'' \emph{Proc. ACM Interact. Mob. Wearable Ubiquitous Technol.},
  vol.~3, no.~3, sep 2019.

\bibitem{JAVADI202048}
S.~H. Javadi and A.~Farina, ``Radar networks: A review of features and
  challenges,'' \emph{Information Fusion}, vol.~61, pp. 48--55, 2020.

\bibitem{8735849}
Z.~Wang, K.~Jiang, Y.~Hou, Z.~Huang, W.~Dou, C.~Zhang, and Y.~Guo, ``A survey
  on {CSI}-based human behavior recognition in through-the-wall scenario,''
  \emph{IEEE Access}, vol.~7, pp. 78\,772--78\,793, 2019.

\end{thebibliography}

%








\end{document}